\documentclass[12pt, draftclsnofoot,onecolumn]{IEEEtran}
\usepackage[dvips]{graphicx}
\usepackage{subfigure}
\usepackage{cite}
\usepackage{float}
\usepackage{amssymb, amsmath}

\begin{document}

\title{GreenDelivery: Proactive Content Caching and Push with Energy-Harvesting-based Small Cells}

\author{Sheng~Zhou, Jie Gong, Zhenyu Zhou, Wei Chen, Zhisheng~Niu
\thanks{Sheng~Zhou, Jie Gong, Wei Chen, and Zhisheng~Niu are with Tsinghua National Laboratory for Information Science and Technology, Dept. of
Electronic Engineering, Tsinghua University, Beijing 100084, China.

Zhenyu Zhou is with the State Key Laboratory of Alternate Electrical Power System with Renewable Energy Sources, School of Electrical and Electronic
Engineering, North China Electric Power University, Beijing, 102206, China.

This work is sponsored in part by the National Science Foundation of China (NSFC) under grant No. 61201191, the National Basic Research Program
of China (973 Program: No. 2012CB316001 and No. 2013CB336600), the National Science Foundation of China (NSFC) under grant No. 61322111, No. 61321061, No. 61401250, and No. 61461136004, and Hitachi Ltd. }}

\maketitle

\begin{abstract}
The explosive growth of mobile multimedia traffic calls for scalable wireless access with high quality of service and low energy cost. Motivated by the emerging energy harvesting communications, and the trend of caching multimedia contents at the access edge and user terminals, we propose a paradigm-shift framework, namely GreenDelivery, enabling efficient content delivery with energy harvesting based small cells. To resolve the two-dimensional randomness of energy harvesting and content request arrivals, proactive caching and push are jointly optimized, with respect to the content popularity distribution and battery states. We thus develop a novel way of understanding the interplay between content and energy over time and space. Case studies are provided to show the substantial reduction of macro BS activities, and thus the related energy consumption from the power grid is reduced. Research issues of the proposed GreenDelivery framework are also discussed.

\end{abstract}

\section{Introduction}
Facing the rapidly growing multimedia traffic over the air and the concern regarding CO2 emissions, it is crucial to innovate green wireless access. In particular, three emerging technologies have been demonstrated as effective, which are, energy harvesting (EH) \cite{Gunduz_2014}, traffic-aware service provisioning \cite{Tango_2011}, and wireless multicasting \cite{Chen_2013}.

More specifically, EH utilizes the energy from natural sources such as solar, wind, and kinetic activities, allowing wireless transmissions to consume less energy \cite{Gong_2013} or no energy \cite{Yener12} from the power grid. With EH based access nodes, such as base stations (BSs) \cite{Gong_2014} and small cells (SCs), a more environmental friendly network can be constructed. Traffic-aware service provisioning was proposed to match the wireless resources to the traffic demands, thereby achieving better energy efficiency (EE). For instance, one can exploit lazy scheduling, i.e., deliver the data with low rate to save power as long as a given deadline is met. Another example is optimizing BS sleeping based on the traffic demands and EH profile \cite{Gong_2014}. Finally, wireless multicast holds the promise of achieving significant EE gain via delivering commonly interested contents to multiple users simultaneously, which avoids duplicated retransmissions of the same content \cite{Chen_2013}\cite{Niu_2008}.

However, there have been some barriers preventing these three methods from being practical. First, exploiting EH is limited by the state of the art readiness for battery capacity. Due to the double randomness and temporal mismatch between energy arrivals and traffic arrivals, a large amount of harvested energy should be stored in batteries, otherwise, energy waste or shortage will occur. Second, the EE gain from on-demand service is also limited because of harsh and stringent QoS requirements of multimedia traffics like video streaming, etc., where many bits should be delivered before an urgent deadline. Finally, in current cellular infrastructures, wireless multicasting can only be enabled if and only if a number of users requires a common content \emph{concurrently}. Otherwise the transmitter has to delay the respond to earlier demands so as to have concurrency, which may severely damage the QoS of the earlier user demands.

To make the above three methods practical, we propose a paradigm-shift framework, namely GreenDelivery, where EH based SCs provide content delivery services. Based on the EH status and content popularity distribution, the SCs proactively cache and push the contents before the actual arrival of user demands. In reward, the time duration in which the desired content can be delivered is greatly extended, so that the delivery can flexibly match the EH process and enjoy low rate transmission. The GreenDelivery framework is built upon the recent trend of providing smart content service with the last-mile wireless access: Contents can be cached at the SCs \cite{Molisch_2013,Molisch_2013b} or relay nodes \cite{Wang_2011}, with proactive caching schemes \cite{Bennis_2014}. The benefits include reducing the core network overhead and enhancing user experiences in terms of delay and rate thanks to shorter access distances. Correspondingly, there have been some initial developments of such technology in commercial products, e.g., HiWiFi\footnote{http://www.hiwifi.com/j2}, Smart WiFi\footnote{http://www.linksys.com/en-us/smartwifi}, with large storage and advanced operating systems capable of running various applications to customize content caching schemes. To further reduce the energy consumption via multicast, proactive push \cite{Podnar_2002} can be introduced on top of caching. The analysis to the network capacity gain provided by proactive push is presented in \cite{Wang_2014}. Practically, incorporating push into current mobile network is supported by the integration of broadcast and communication network \cite{Niu_2008}. Moreover, as the EH technology has been applied to access nodes \cite{Gong_2014}, researchers are considering using such EH based SCs to cache contents \cite{GreenCache_2013}, getting the best of their deployment flexibility and low CO2 emission.

In this article we provide a more general picture with GreenDelivery framework, with the joint design of EH, push, and caching to provide two-fold benefits of QoS and greenness. Enabling content caching nodes with push capabilities, specifically for EH based SCs, can well match the multimedia traffic with random energy arrivals. The benefit will be reflected in the reduction of macro BS activities and thus the reduced energy consumption from the power grid. The next section will overview the framework, the intuitive ideas behind it, and its benefits. In subsequent sections, two case studies are illustrated, and related research challenges are discussed.	

\section{GreenDelivery: The Framework}

\begin{figure}[t]
\centering
\includegraphics[width=5in]{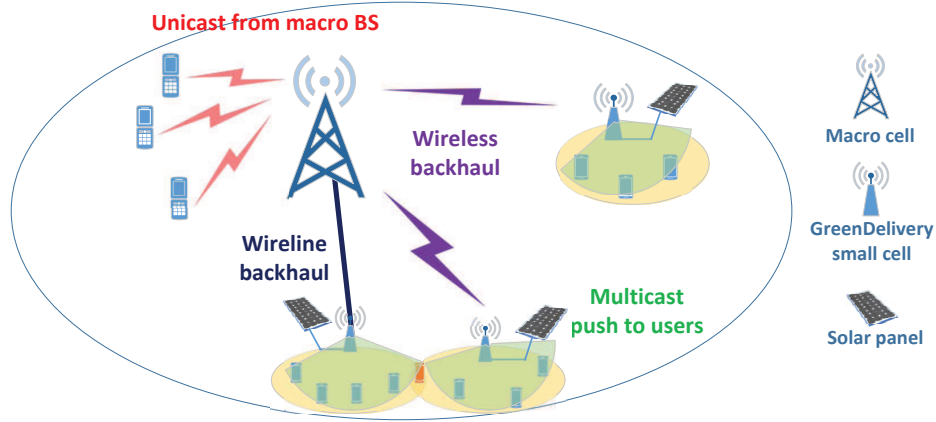}
\caption{The concept of GreenDelivery. The small cells are energy harvesting based while the macro BS is grid powered.} \label{fig:concept}
\end{figure}

The concept of GreenDelivery is illustrated in Fig.~\ref{fig:concept}, where multiple GreenDelivery SCs cache popular contents and push them to users proactively.  The design \emph{objective} of GreenDelivery framework is to minimize the number of user requests handled by the macro BS. The intuition of such metric is two-fold: First is energy saving. As the macro BS generally connects to the power grid, minimizing the activity of the macro BS reduces the grid power consumption, while the renewable energy used to power the GreenDelivery SCs can be regarded as free. Note that the backhaul link from the BS to SCs is generally good, and can enjoy low power transmission. The second consideration is the user quality of service (QoS). For those contents already pushed to users, users can get the contents with zero delay. Even if for those requested contents that have not been pushed, as SCs are closer to users, unicast from the SC provides higher transmission rate and thus guarantees shorter delay.

Specifically, EH technology provides renewable energy for GreenDelivery SCs to:
\begin{itemize}
\item Fetch contents from macro BS via the backhaul link. Since these SCs are EH based, it is reasonable that they may only have wireless backhaul. As a result, the energy of fetch is not negligible, and accounts among major consumption portions of the harvested energy. Note that the wireline backhaul, as shown in Fig.~\ref{fig:concept}, can also be considered as an option with less energy consumption but higher deployment cost.

\item Cache the fetched content. The energy consumption depends on the storage method, and the content storage volume. For GreenDelivery SCs serving limited number of users, the contents can be stored locally in the SC hardware, with negligible energy consumption for caching. When the cache size is large, additional hardware, like cache server \cite{GreenCache_2013}, is required, and the energy consumption cannot be ignored.

\item Push the contents to users before the user could potentially request it. Once contents have been fetched and cached at the GreenDelivery SC, they are selected to be pushed to users depending on their popularity and the battery status. As shown in Fig.~\ref{fig:concept}, not only the users associated to the SC, but also those (as the handset in red) in the overlapping coverage of multiple SCs can enjoy the push service, from one or more SCs respectively. In other words, the caching and push can be coordinated among multiple SCs, and one example of caching coordination can be found in \cite{Molisch_2013b}. Note that for commonly interested contents, multicast/broadcast is performed, while for private contents, unicast push is performed.

\item Unicast the contents to users upon request. Users may request to its associated SC for some content before it is pushed. If the SC has the content fetched and the battery has enough energy, it will unicast the content to the user upon request. Pushing the requested content can also be considered, but if the requested content is private or not popular, it is not beneficial considering the limited storage on user terminals.

\end{itemize}

On the user side, if the upper layer application requests some content, the user will check its local storage first to see if the content has already been pushed. If not, it will request over the air to its associated SC. Note that in this paper, the request is still counted even if it is satisfied by the local storage. If the SC is not able to handle the request, the macro BS takes over and unicasts the content to the user.

The push mechanism can be realized by the existing broadcasting protocols without additional signaling overhead. Options include multimedia broadcast multicast services (MBMSs) proposed by 3GPP, or its new version called broadcast and multicast service (BCMCS) \cite{Wang_2014}. The coordination of such broadcast channel falls into the category called integrated communication and broadcast networks (ICBN) \cite{Niu_2008}. If specified broadcast channel does not exist, the GreenDelivery SC can reuse the unicast channel, but in this case the users besides the default unicast receiver should be notified to receive via dedicated signaling. The signaling can be conveyed via the downlink control channel. Note that SCs should have the popularity distribution of the contents, which can be updated by the macro BS periodically. This overhead is proportional to the update rate of the user interests, which is generally slow compared to the time scale of data transmission and EH.

\subsection{Exploiting the Content and Energy Timeliness}

The key idea of GreenDelivery is to exploit the timeliness of the contents and energy via intelligent caching and push, so that to match random energy arrivals and user requests over time and space. The timeliness of the contents corresponds to their popularity and life span. The contents can only be interested to users for a finite period of time, and the popularity ranking of the contents may change over time. The timeliness of the harvested energy comes with the causality of energy usage and the limited battery capacity. The causality means that the harvested energy cannot be used before its arrival. Moreover, the limited battery capacity brings constraints on the delay of using such energy. In other words, if the arrived energy is not used timely, newly arrived energy will be wasted when the battery is full. The two-fold randomness poses challenges in delivering contents efficiently. To solve this, based on the popularity and life span of the contents, together with the battery status, the GreenDelivery SC proactively fetches and caches the popular contents and then pushes them to users. In this way, the delay constraint of using the harvested energy is resolved, since the energy is used to provide stored contents at the users via push without waiting for the request. Or the harvested energy can be regarded as if it is transferred from the time when the content is pushed, into the \emph{future} when the user actually requests the content. This is another way of information and power transfer over the hyper dimension of space (small cell to users) and time (present to the future), which is different from the information and power transfer over space only \cite{Zhou_2013}.

\begin{figure}[t]
\centering
\includegraphics[width=5in]{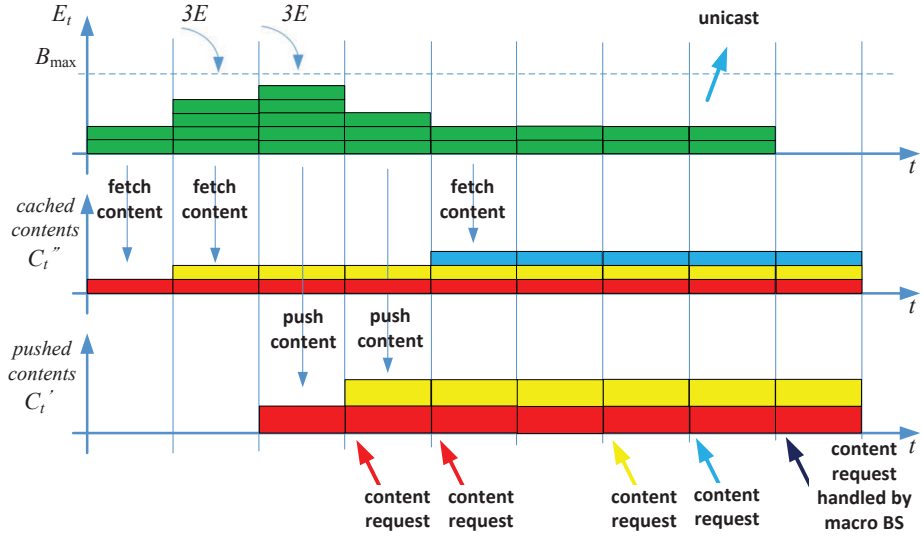}
\caption{The behavior of a GreenDelivery SC: An example.} \label{fig:timeline}
\end{figure}

An illustrative example of the aforementioned idea is shown in Fig.~\ref{fig:timeline}. The time horizon is divided into equal-length periods, which can be regarded as the broadcast frames on the broadcast channels in MBMS, and user requests arrived during some period are batched and responded at the beginning of the next period. Energy is harvested and stored in the battery of the SC at the beginning of each period. In this example the SC can fetch or transmit (push or unicast) at most one content in each period. The set of cached contents as of period $t$ is denoted by $\mathcal{C}_t''$, and the set of pushed contents as of period $t$ is denoted by $\mathcal{C}_t'$. Assume the length of the contents is the same, and the height of the contents in the figure represents the energy used to fetch or push it, i.e., unit energy $E$ for fetching and caching a content, and 2$E$ for pushing a content. As shown in Fig.~\ref{fig:timeline}, the energy arrives in the second and third periods, the SC utilizes these energy (including the initial energy in the battery) to fetch and cache two contents. The SC then pushes the most popular content (the red one) in its cache to its users. Consequently, at the fourth and fifth period, two requests for the red content arrive, and since it is pushed, the requests are instantaneously satisfied locally at user terminals. In the mean time, the SC can push and fetch more contents, satisfying the requests in the seventh and eighth periods. Note that the request for the blue content is served with unicast from the SC as it is not pushed yet. In the last period, since the SC is running out of energy, the user request is responded by the macro BS. In the example, without proactive caching and push, the four requests except the last one requires 11$E$ ($2E \times 4 = 8E$ for content unicast, and $E \times 3 = 3E$ for fetching the three contents), which cannot be satisfied since: The total energy budget is only $8E$; The battery capacity is $6E$, which means that if the energy is not used proactively, $2E$ of the harvested energy will be wasted.

\subsection{Benefits of GreenDelivery}
First, the temporal mismatch of content requests and energy arrivals can be resolved. Since the contents are cached, the SC may carry out content delivery whenever there is enough harvested energy in the battery. On the other hand, since the harvested energy can be effectively and timely used, energy waste due to battery overflow can be avoided.

Second, the SC push can greatly benefit from low rate transmission, since there is no urgent delay constraint for proactive push, and thus the SC is allowed to transmit with reasonable low power. Note that there exists a non-zero EE-optimal transmission time to balance the transmission and circuit power when holistic power model is taken into account. Hence, the push holds the promise of achieving this EE-optimal transmission time in practice.

Finally, the joint use of push and caching enables more opportunities of wireless multicasting. During proactive push, all users who are potentially interested in these contents may overhear and decode the signal. In this case, wireless multicasting will not delay the service to any user because it is very flexible to align the time of push to different users. The only cost to be paid is the storage resource for caching, the price of which is dramatically dropping nowadays \cite{Molisch_2013}.

\section{Case Study: Content Caching and Push under Dynamic Energy Arrivals}

\begin{figure}[t]
\centering
\includegraphics[width=5in]{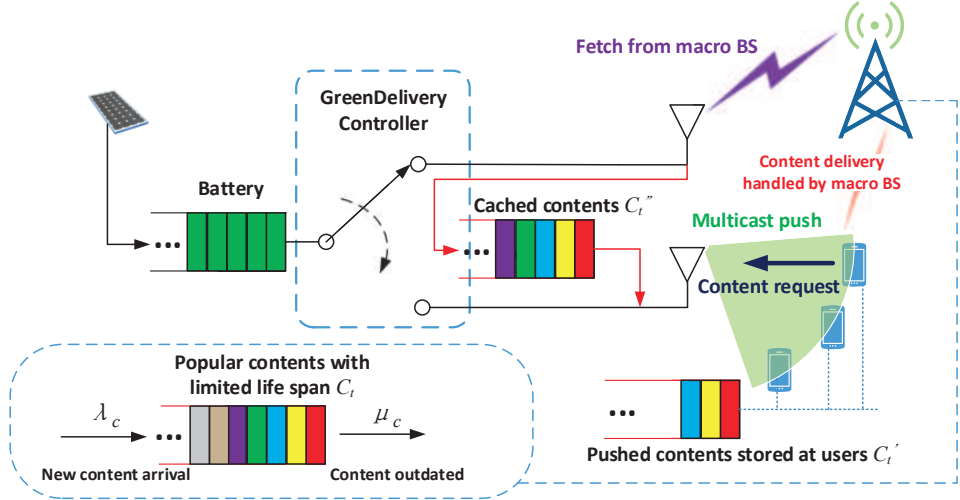}
\caption{Block diagram of a GreenDelivery small cell.} \label{fig:example}
\end{figure}

In this section, we express our idea through two case study examples. The considered model is presented in Fig.~\ref{fig:example}, where one EH-based GreenDelivery SC is illustrated. In each time period, the SC harvests a random amount of energy units, and stores it in a battery with finite capacity $E_{\mathrm{max}}$. Assume one or more contents can be fetched in a period, with energy consumption $E_F$, and one content can be transmitted, through either push or unicast, in a period with $E_P$ units of energy. For example in Fig.~\ref{fig:timeline}, $E_F = E$ and $E_P = 2E$.
Note here we simplify the channel to be static so that all transmissions are assumed successful. Specifically, the channel from the macro BS to SC has slight impact on the energy for fetch $E_F$ as it is the receiving energy. The channels from SC to users are assumed to be static and identical, so that $E_P$ amount of push energy can always guarantee successful delivery.

The active content set $\mathcal{C}_t$ is time-varying, i.e., new content comes into play over time and contents can also be outdated. This property is described as a birth-death process with birth rate $\lambda_c$ and death rate $\mu_c$ (in the unit of per period), and the number of active contents at time period  $t$ is given by $|\mathcal{C}_t| = m$. The popularity of a content is defined as the probability that a user request corresponds to this content, denoted by $f_m^i$, which follows Zipf distribution \cite{Molisch_2013}. By ranking the $m$ contents with descending popularity, the popularity of the $i$-th ranked content is $f_m^i = 1/(i^v\sum_{j=1}^m1/j^v)$, where $v>0$ is the Zipf parameter, and larger $v$ means that fewer contents account for the most popular ones.
The user request arrives at the beginning of a period with probability $p_r$, and we count both kinds of requests: those satisfied by the user local storage filled by proactive push, and the requests served over the air from the SC. In addition, when a content gets outdated and departs from the active content set $\mathcal{C}_t$, it will be removed from both $\mathcal{C}_t''$ and $\mathcal{C}_t'$ (if it is fetched and pushed to users), and the corresponding storage space is released.

At the beginning of each period, based on the current system state, including the active content set $\mathcal{C}_t$, the pushed content set $\mathcal{C}_t'$, the cached content set $\mathcal{C}_t''$ and the amount of energy units in the battery, the GreenDelivery Scheduler makes the action decision. The action set includes: fetch a content for caching; push a content; unicast a required content in $\mathcal{C}_t''$; do nothing. When the SC decides to do nothing, the user request, if arrives, will be handled by the macro BS. As explained, our policy design objective is to minimize the ratio of user requests handled by the macro BS, denoted by $\eta$, subject to the energy causality constraint, i.e., the energy can not be used before its arrival.  In what follows, we will first investigate the push behavior of GreenDelivery SC, and then both fetch (for caching) and push will be considered.

\subsection{Energy-Harvesting-based Proactive Push}

To reflect the gain provided by proactive push, we first consider push only. Assume $\mathcal{C}_t'' = \mathcal{C}_t$, which corresponds to the case that the SC can get the content instantaneously via high-speed backhaul when it needs to push or unicast it. Therefore the energy consumption of fetch and caching is ignored.
When the energy in the battery is sufficient and the SC decides to push, the most popular content in $\mathcal{C}_t''$ is pushed to users. We assume that the storage space of users are large enough to store the contents in set $\mathcal{C}_t'$.

We consider a simple energy-aware push scheme to see how proactive push can reduce the probability of handling user requests by the macro BS. Specifically, if there is no user request \emph{over the air} in current period and the battery energy is sufficient for pushing a content, the SC will push the most popular content in $\mathcal{C}_t''$ which has not been pushed. Otherwise if a user requests a content, a unicast will be performed given the battery energy is sufficient to push a content. The user request is handled by the macro BS when the energy in the SC battery is not enough, and in this case the SC will do nothing in current period.

\begin{figure}[t]
\centerline{\subfigure[$E_{\mathrm{max}} = 10, v=1$.]{\includegraphics[width=2.2in]{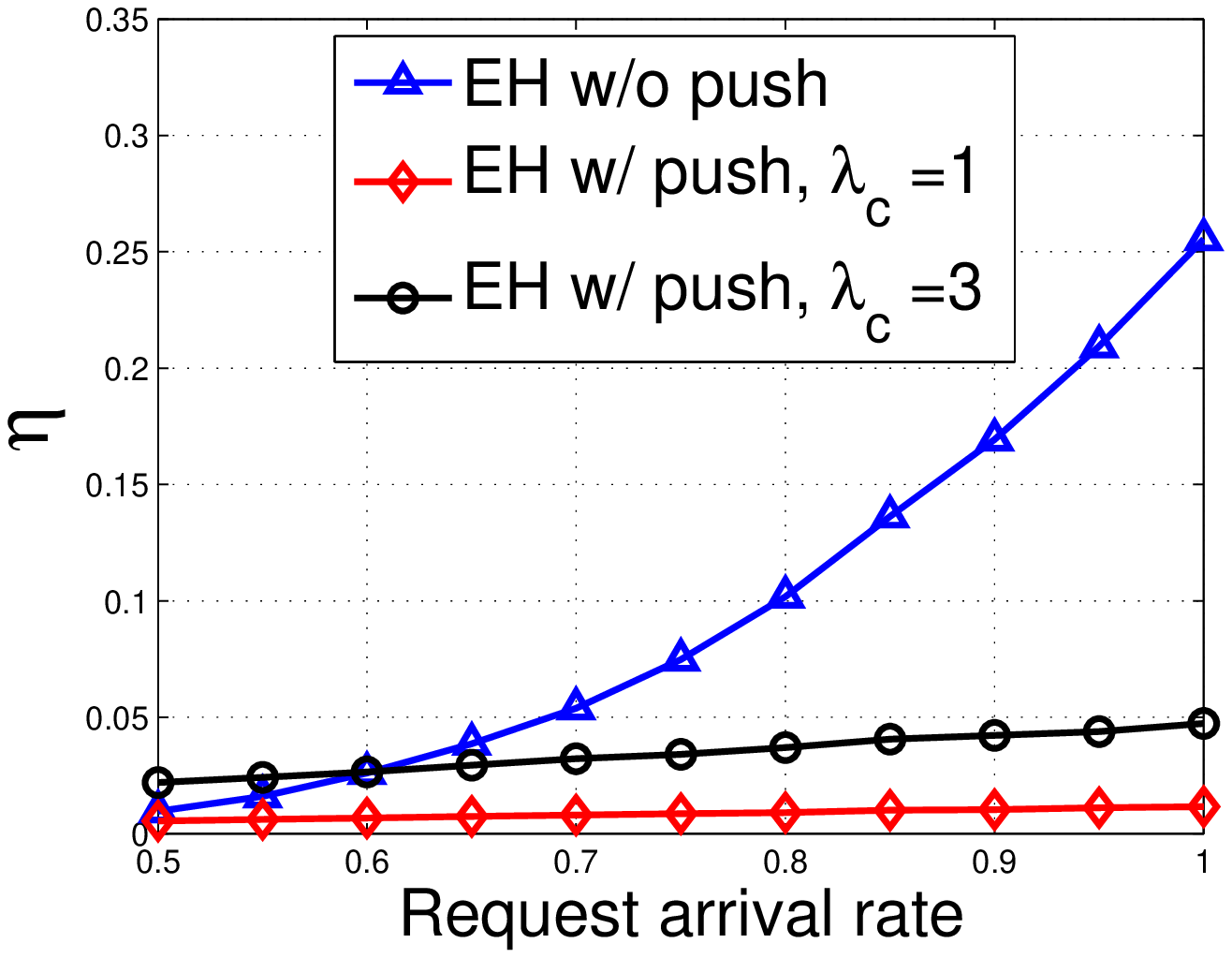}
\label{fig:push02}}
 \subfigure[$p_r = 0.75, v=1$.]{\includegraphics[width=2.2in]{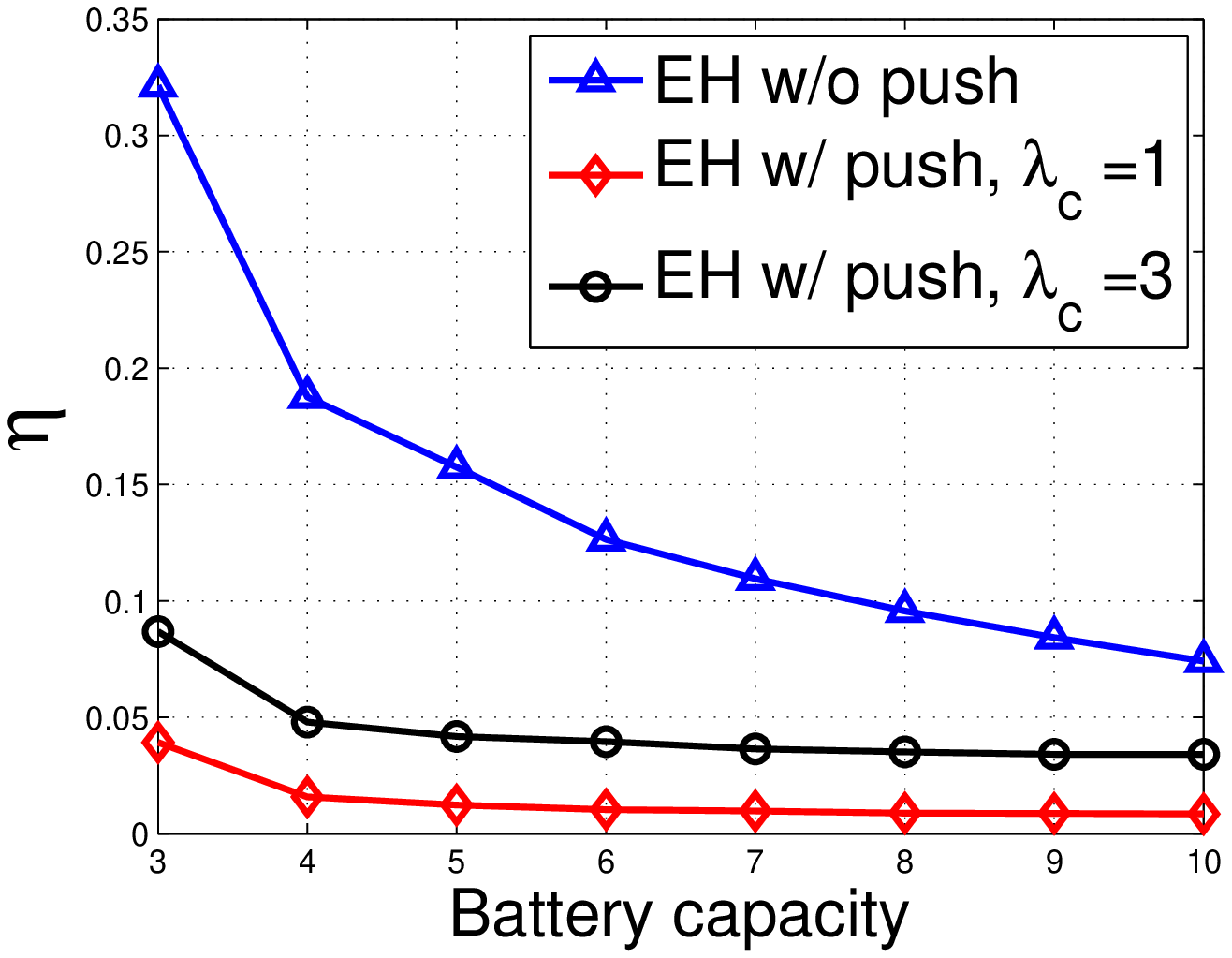}
\label{fig:pushvsEmax02}}
 \subfigure[$E_{\mathrm{max}} = 10, p_r=0.75$.]{\includegraphics[width=2.2in]{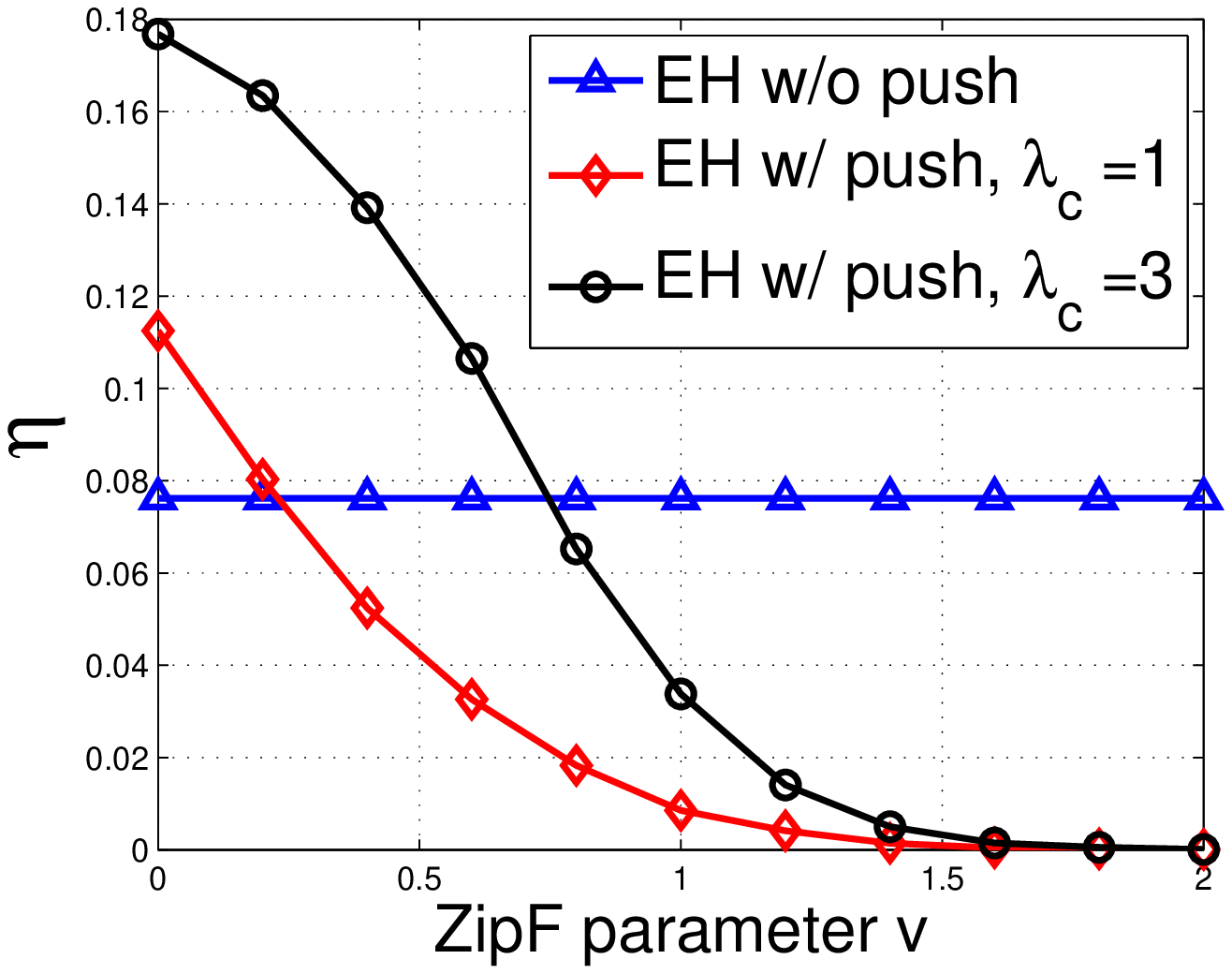}
\label{fig:pushvsv02}}} \caption{Evaluations of the ratio of the requests handled by the macro BS, with proactive push, $E_P = 2, E_H = 3$, where $\eta$ denotes the the ratio of user requests handled by the macro BS.}
\label{fig:push}
\end{figure}

Suppose the energy arrival follows Bernoulli distribution, i.e., at the beginning of each period, the system can harvest $E_H$ units of energy (we omit the notation $E$ of unit energy hereafter) with probability $p$. We set $p=0.5$, $\mu_c=1\times 10^{-3}$ per period, $v = 1$, $E_P = 2$ and $E_H=3$. The ratio of requests handled by the macro BS is shown in Fig.~\ref{fig:push}. To compare, the policy without proactive push serves as the baseline, where the SC unicasts a required content as long as there is enough energy in the battery, or if no user request arrives the SC does nothing. It can be seen from Fig.~\ref{fig:push02} that in the baseline scheme, the probability that the SC cannot provide service increases as the request arrival rate $p_r$ increases. While on the other hand, proactive push keeps such probability low and stable, i.e., almost irrelevant to $p_r$, hence greatly reduces the burden of the macro BS. One can also note that when $p_r$ is low and the content refreshing rate is high, i.e., $\lambda_c=3$ per period, proactive push does not bring any performance gain, because in this case user requests are diverse over different contents and a pushed content has low probability to be requested by multiple users.

We also consider the influence of the battery capacity $E_{\max}$, since one major benefit of GreenDelivery is to solve the energy availability issue with limited battery. As shown in Fig.~\ref{fig:pushvsEmax02}, the ratio of requests handled by the macro BS decreases as $E_{\max}$ increases. Compared to the baseline, the reduction of the ratio is significant. In other words, to achieve the same performance, the required battery capacity with push is smaller than that without push.

In Fig.~\ref{fig:pushvsv02}, the impact of the content popularity distribution is depicted, where the Zipf distribution parameter $v$ is varying from $0$ to 2, i.e., from a uniform distribution to a more skewed one. When the contents are uniformly distributed, it is better not to use proactive push, while the gain of proactive push increases with more skewed content distribution and lower content refreshing rate.

\subsection{Energy-Harvesting-based Caching and Push}
\begin{figure}[t]
\centerline{\subfigure[$E_{\mathrm{max}} = 10, v=1$.]{\includegraphics[width=2.2in]{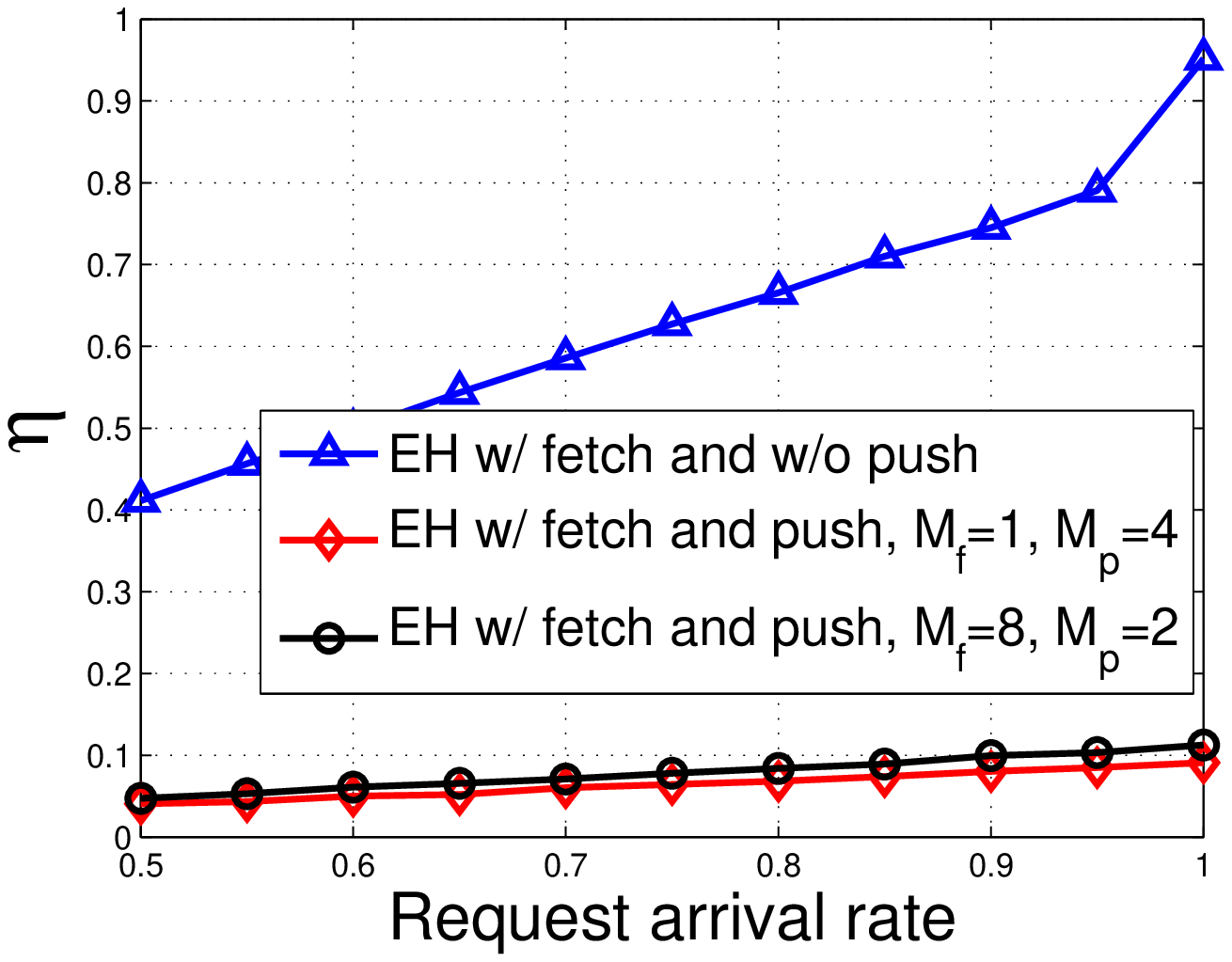}
\label{fig:pushfetchvcpr}}
 \subfigure[$p_r = 0.75, v=1$.]{\includegraphics[width=2.2in]{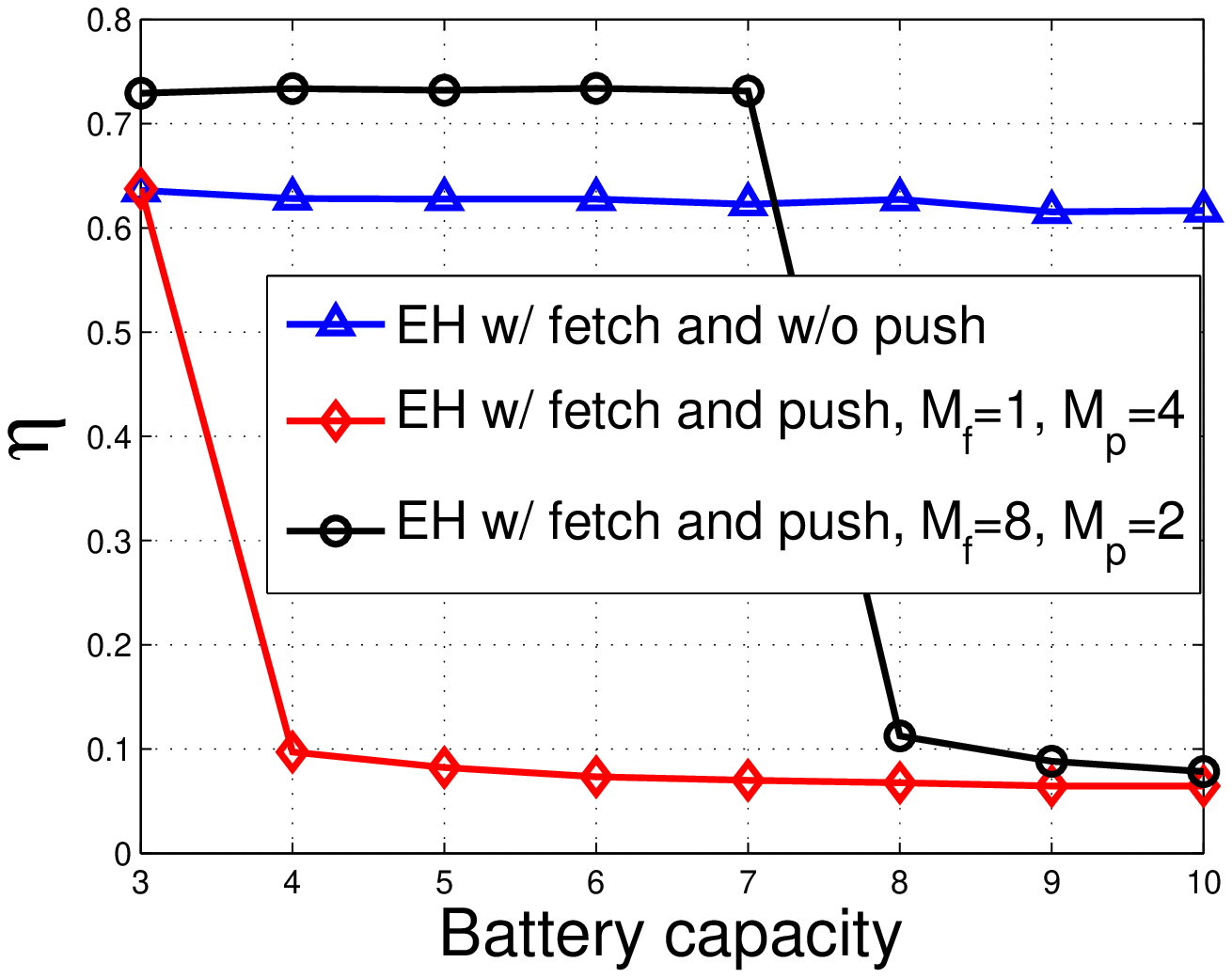}
\label{fig:pushfetchvsEmax}}
 \subfigure[$E_{\mathrm{max}} = 10, p_r=0.75$.]{\includegraphics[width=2.2in]{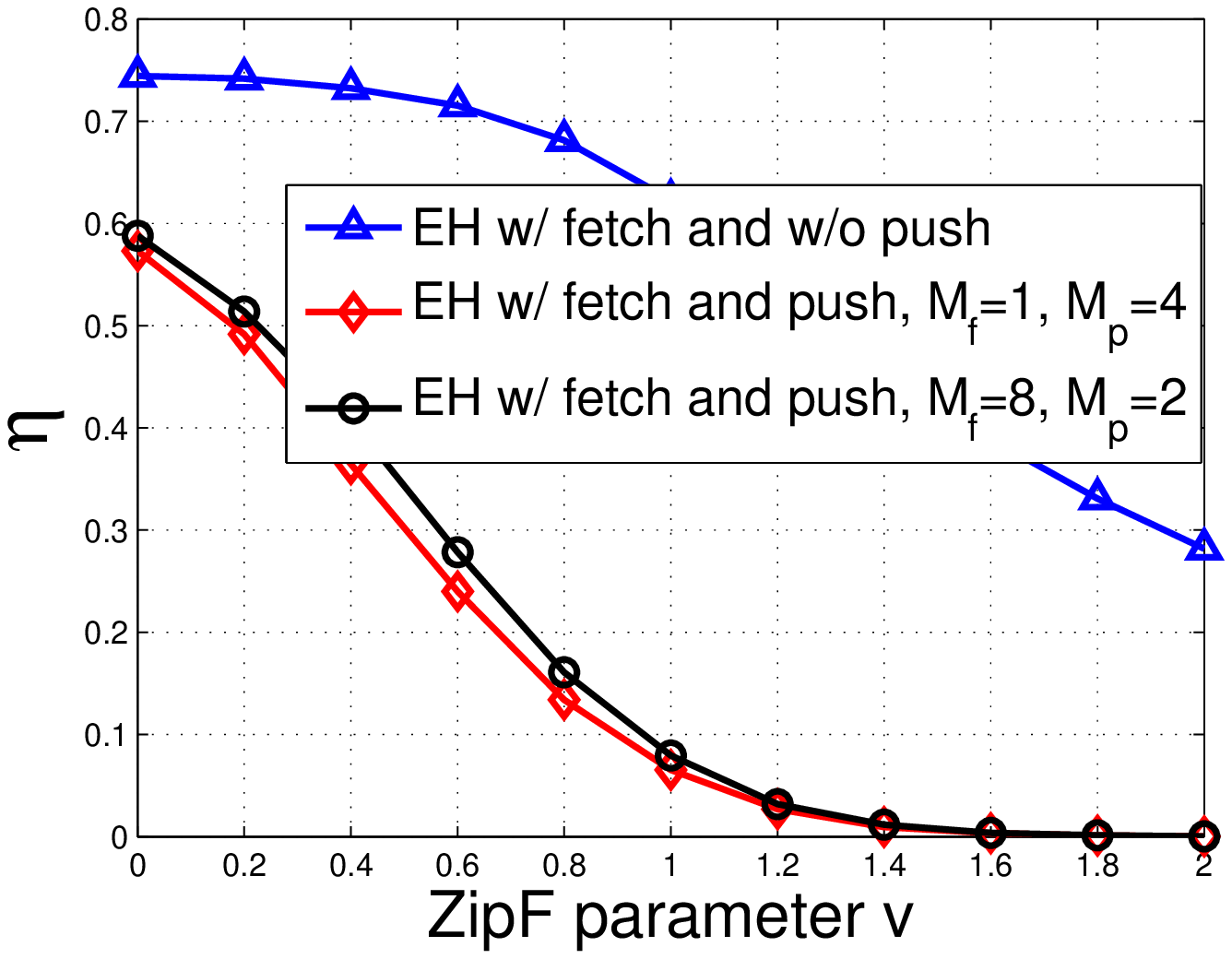}
\label{fig:pushfetchvsv}}} \caption{Evaluations of the ratio of the requests handled by the macro BS, with proactive fetch for caching and push, $E_F=1, E_P = 2, E_H = 3$, where $\eta$ denotes the the ratio of user requests handled by the macro BS.}
\label{fig:pushfetch}
\end{figure}

We then take into account the cost of fetching the contents to cache at the GreenDelivery SC. Initially the set of contents $\mathcal{C}_t$ is not available at the SC, and the SC has to firstly fetch the contents from the macro BS via the backhaul. It is reasonable to assume that the energy for fetching a content is less than that for pushing a content, and SC can possibly fetch multiple contents in one period. A threshold based fetch and push policy is proposed.
If the ratio of $|\mathcal{C}_t'|$ to $|\mathcal{C}_t''|$ is higher than the ratio of $|\mathcal{C}_t''|$ to $|\mathcal{C}_t|$, this means that the number of cached contents is relatively small and the SC needs to fetch more contents to avoid requests served by the macro BS. Then if the number of energy units in the battery is no less than a given threshold $M_f$, the SC fetches at most $K$ contents and consume $E_F$ units of energy. There is another threshold $M_p$ for push. If the ratio of $|\mathcal{C}_t'|$ to $|\mathcal{C}_t''|$ is lower than that of $|\mathcal{C}_t''|$ to $|\mathcal{C}_t|$, the cached contents in the SC need to be pushed to reduce the possible unicast events. In this case, the SC will push the most popular content in $\mathcal{C}_t''$ that has not been pushed, given the battery energy is no less than $M_p$. Finally, the user request is handled by the macro BS when the energy in the SC battery is not enough or the content has not been fetched, and in this case the SC will do nothing in current period. Intuitively, larger thresholds $M_f$ and $M_p$ decrease the probability of battery outage when a request arrives, but on the other hand reduce the chances of push or fetch.

We set $\lambda_c = 1$ per period, $K = 3, E_F = 1, E_{\mathrm{max}} = 10$, and other parameters are the same as those in the previous section. In Fig.~\ref{fig:pushfetch}, we compare the proposed algorithm with the case without proactive push. Similar to Fig.~\ref{fig:push}, it shows that push can significantly improve the performance, and this also confirms the necessity of having push in GreenDelivery. The results also indicate that the thresholds should be carefully selected regarding the system parameters, especially the battery capacity as shown in Fig.~\ref{fig:pushfetchvsEmax}. From the figures, it is conjectured that having more aggressive fetch and caching (with smaller $M_f$) provides better performance, as contents to be pushed should already been cached. But one should also note that using too much energy for fetch and caching leaves less energy for push, so that these two sides of activities should be balanced.

\section{Research Challenges}
Several research challenges for releasing the benefits of the GreenDelivery framework are discussed as follows.
\subsection{Intelligent Push under Random Energy Arrivals, Finite Battery and Fading Channel}
In practice, the energy arrival and user requests can not be precisely predicted. Therefore, online algorithms in charge of intelligent caching and push are required, with low computational complexity but close-to-optimal performance. Intuitively, more contents should be fetched and pushed if the energy arrival is sufficient and the battery is full, in order to avoid the battery overflow. On the other hand, when the harvested energy is not enough, it is better to reserve energy to handle randomly arrived user requests. Moreover, as in reality the channel fading brings another dimension of randomness, the energy used for push is also varying. As a result, the algorithm design is not straightforward due to the three-fold randomness of EH process, user request arrivals, and channel fading.

\subsection{Learning and Prediction of the Popularity and Energy Arrival Statistics}
If the statistical information of either content popularity or energy arrival process is unknown, the online algorithms should be able to learn and predict content popularity and energy arrival statistics. Due to the huge data volume and content variety, emerging big data learning technologies may be employed to provide both short-term and long-term popularity prediction. On the other hand, solar radiation models or wind speed models have been systematically studied for several decades. However, it is still an open problem to effectively combine popularity prediction and energy arrival prediction together into the resource management for GreenDelivery SCs.

\subsection{Trade-off between Benefits of Push and Content Storage Cost}
In this article, we mainly consider the energy of fetch and push, while caching itself also introduces additional costs including both energy consumption and storage occupancy. Large cache enhances the capacity of caching and push, but on the other hand is expensive and energy consuming. Thus the challenging problem is how to model and quantize the content storage cost and how to achieve the optimal trade-off between the caching cost and the benefit of push. One could also improve the trade-off relation by caching and push the prefix of the multimedia contents rather than the whole, especially for videos so that the initial play out delay can still be reduced.

\subsection{Cooperation among Multiple GreenDelivery Small Cells}
GreenDelivery SCs can be densely deployed. Hence, it is possible to have cooperation and interaction among adjacent SCs to jointly optimize both customer-level QoS and system-level performance. For example, some SCs with larger request rate but low energy harvesting rate or small battery capacity need the help from neighboring SCs. Multiple SCs can also form a cluster of coordinated transmission to combat channel fading. However, as the contents cached in different SCs are generally not the same, they may need to re-fetch the contents before push, which costs additional energy. Therefore, adjacent SCs should coordinate the re-fetch and the push behaviors to efficiently handle the content delivery, which poses design challenges especially for large-scale heterogeneous network.

\section{Conclusion and Outlook}
In this article, GreenDelivery as a new access network framework is proposed to enable efficient content delivery via EH based SCs. Exploiting the content popularity information and battery status, proactive fetch/caching and push are implemented to match the random energy arrival and user content requests, and to provide more multicast opportunities. In this way, the limited harvested energy is wisely used, and the transmission cost of macro BSs is substantially reduced, which is illustrated via our case studies. We believe the idea of GreenDelivery is promising for delivering multimedia contents with densely deployed EH-based SCs, enjoying their deployment flexibility and energy scalability. GreenDelivery also motivates some future research directions including online policies for joint fetch-caching-push, learning of the EH and content statistics, and cooperation among multiple GreenDelivery SCs.


\vspace*{-2\baselineskip}

\begin{IEEEbiographynophoto}{\bf Sheng Zhou} [M] (sheng.zhou@tsinghua.edu.cn) received his B.S. and Ph.D. degrees in Electronic Engineering from Tsinghua University, China, in 2005 and 2011, respectively. He is currently an assistant professor of Electronic Engineering Department, Tsinghua University. From January to June 2010, he was a visiting student at Wireless System Lab, Electrical Engineering Department, Stanford University, CA, USA. His research interests include cross-layer design for multiple antenna systems, cooperative transmission in cellular systems, and green wireless communications.
\end{IEEEbiographynophoto}

\vspace*{-2\baselineskip}

\begin{IEEEbiographynophoto}{\bf Jie Gong} [M] (gongj13@tsinghua.edu.cn) received the B.S. and Ph.D. degrees from Tsinghua University, Beijing, China, in 2008 and 2013, respectively. He is currently a PostDoc with Tsinghua University. From July 2012 to January 2013, he visited the University of Edinburgh, UK. His research interests include base station cooperation in cellular networks, energy harvesting, and green communications.
\end{IEEEbiographynophoto}

\vspace*{-2\baselineskip}

\begin{IEEEbiographynophoto}{\bf Zhenyu Zhou} [M] (zhenyu\_zhou@fuji.waseda.jp) received his M.E. and Ph.D degree from Waseda University, Tokyo, Japan in 2008 and 2011 respectively. From April 2012 to March 2013, he was the chief researcher at Department of Technology, KDDI, Tokyo, Japan. From March 2013 to now, he is an Associate Professor at School of Electrical and Electronic Engineering, North China Electric Power University. His research interests include energy-efficient wireless communications, and smart grid communications.
\end{IEEEbiographynophoto}

\vspace*{-2\baselineskip}

\begin{IEEEbiographynophoto}{\bf Wei Chen} [SM] (wchen@tsinghua.edu.cn) received his BS degree in Operations Research and PhD degree in Electronic Engineering (both with the highest honors) from Tsinghua University in 2002, and 2007. He is a full professor and deputy head of Electronic Engineering Department, Tsinghua University, as well as a National 973 Youth Project chief scientist and a winner of National May 1st Medal. He received IEEE Comsoc APB Best Young Researcher Award and IEEE Marconi Prize Paper Award.
\end{IEEEbiographynophoto}

\vspace*{-2\baselineskip}

\begin{IEEEbiographynophoto}{\bf Zhisheng Niu} [F] (niuzhs@tsinghua.edu.cn) graduated from Beijing Jiaotong University, China, in 1985, and got his M.E. and D.E. degrees from Toyohashi University of Technology, Japan, in 1989 and 1992, respectively.  During 1992-94, he worked for Fujitsu Laboratories Ltd., Japan, and in 1994 joined with Tsinghua University, Beijing, China, where he is now a professor at the Department of Electronic Engineering and deputy dean of the School of Information Science and Technology.  He is also a guest chair professor of Shandong University, China.  His major research interests include queueing theory, traffic engineering, mobile Internet, radio resource management of wireless networks, and green communication and networks.  He is now a fellow of both IEEE and IEICE, a distinguished lecturer (2012-15) and Chair of Emerging Technology Committee (2014-15) of IEEE Communication Society, and a distinguished lecturer (2014-16) of IEEE Vehicular Technologies Society.
\end{IEEEbiographynophoto}

\end{document}